\documentstyle[12pt]{article}
\begin{document}

\title{Solution of Stationary and Axisymmetric Problem in General
Relativity.Nontraditional Approach}

\author{G. H. Haroutyunian}

\date{\today}

\maketitle

\begin{abstract}
Physically admissible choice of the "essential" coordinates identified
with components of the metric tensor and co-moving frame of reference
reduced to the formulation of the stationary axisymmetric GR
problem. Such nontraditional approach allows to obtain new 4-parametric
vacuum solution. In special static case this solution becomes the set
well-known GR exact solutions for degenerate vacuum gravitational fields,
in some special stationary case, after transformations coincides with
Kerr solution. In other special stationary case the founded solution turn
into modified NUT solution.
\end{abstract}

\vspace{3cm}

\section{ Coordinates, frame of reference and equations of problem.}
Stationary and axisymmetric gravitational fields accept the group of
motion with two linearly independent and commuting Killing vectors $\xi$
and $\eta$. According to symmetry of the problem let us to choose
coordinates

\begin{eqnarray}
\label{1}
x^a= \left\{t,\varphi \right\},\qquad a,b, \dots=0,3
\end{eqnarray}

Consequently

\begin{eqnarray}
\label{2}
\begin{array}{ccc}
\xi &=& \xi^{\mu}\frac{\partial}{\partial x^{\mu}}, \,\,\,
\xi^{\mu}=\delta^{\mu}_t,
\\ [4mm]
\eta &=& \eta^{\nu}\frac{\partial}{\partial x^{\nu}}, \,\,\,
\eta^{\nu}=\delta^{\nu}_{\varphi}.
\end{array}
\end{eqnarray}
(Here: $\varphi$ is azimuths angle).

The space-time metric corresponding such gravitational fields must be
forminvariant with respect to transformations

\begin{eqnarray}
\label{3}
x^a= \alpha^a_b {\tilde x}^b, \,\,\, \alpha^a_b = Const,
\end{eqnarray}

and besides allowed coordinates mapping
\begin{eqnarray}
\label{4}
x^A = x^A ({\tilde x}^B),\,\,\, A,B,\dots = 1,2
\end{eqnarray}
on the two dimensional surface $x^1x^2$.The choice of the "essential"
coordinates  $x^1$ and $x^2$ noticeable simplifies solution of problems.
In particular, for vacuum problems, the Weyl canonical coordinates prove
to be preferred, but for internal problems ones give off the own
character.
Some from several internal solutions for rigidly rotating incompressible
fluid founded by Wahlquist [1] which used the generalized ellipsoidal
coordinates. Bonanos and Sklavenites [2,3] suggested to use the
arbitrariness
contained in equation (4) that coordinates $x^1$ and $x^2$ to identify
with components of the metric tensor. Let us followed [2,3] to choose
\begin{eqnarray}
\label{5}
x^1 \equiv \rho = g_{33},\,\,\,  x^2 \equiv \Phi = g_{00}
\end{eqnarray}

In this case the space-time metric is
\begin{eqnarray}
\begin{array}{ccc}
\label{6}
ds^2 = \Phi^2 {(dt-qd \varphi)}^2-e^{2\alpha (\rho, \Phi)}{(d \rho-ld
\Phi)}^2-
e^{2 \beta {( \rho, \Phi )}}d \Phi^2 - \rho ^2d \varphi ^2
\\ [4mm]
q = q( \rho, \varphi), \,\,\, l=l( \rho, \varphi ).
\end{array}
\end{eqnarray}

Such choice efficiently natural, since for weak gravitational fields
$\Phi \approx 1+2 \psi$ be found proportional the Newton potential $\psi$,
and $\rho$ coincides with the cylindrical coordinate of the flat space.
The invariance with respect to inversion $(t,\varphi) \longrightarrow
(-t,-\varphi)$ means that the motion of the gravitational field source
is the pure rotation. In other words the space-time with metrics (6)
was generated the rotating body of which the 4-velocity vector is

$$ u^ \mu = \left \{ u^t,0,0,u^{\varphi} \right \}. $$

If used the type (3) transformations, namely $\varphi \rightarrow
\varphi + \Omega t$,
then we can turn $u^\varphi = 0$, choosing co-moving coordinates
\begin{eqnarray}
\label{7}
u^ \mu = \left \{ \frac{1}{\Phi},0,0,0 \right \}.
\end{eqnarray}

Let assume rigidly rotation with $\Omega=d \varphi /dt$, also suppose that
the matter of the considered configurations is the perfect fluid with

$$ T^{\mu}_{\nu}=(\varepsilon + \sl{P})u_{\mu}u^{\nu}-
{\sl{P}}\delta^{\mu}_{\nu} $$
and the equation of state $\sl{P}=\sl{P}(\varepsilon)$. (Here
$\varepsilon$ - energy
density, $\sl{P}$ - pressure of matter).

Well-known equation $\nabla_{\nu} T_{\mu}^{\nu}=0$ in the metric (6)
and co-moving frame of reference give us
\begin{eqnarray}
\label{8}
\frac{\partial{\sl{P}}}{\partial{\rho}}=0, \,\,\,
\frac{\partial{\sl{P}}}{\partial{\Phi}}=-(\varepsilon + \sl{P})/{\Phi}
\end{eqnarray}

This mean $\sl{P}=\sl{P}(\Phi)$ and $\varepsilon=\varepsilon(\Phi)$.
In other words isobaric $(\sl{P} = Const)$ surfaces coincides with
one of the constant values $\Phi = \Phi_0$. Let us take
constant the value of the free falling test-body acceleration
\begin{eqnarray}
\label{9}
a=\sqrt{F_{\mu}F^{\mu}}=e^{-\beta}/{\Phi},\,\,\,
F_{\mu}=2u^{\alpha} u_{[{\mu},{\alpha}]}
\end{eqnarray}
on the surface $\Phi = \Phi_0$ then we can assume

\begin{eqnarray}
\label{10}
\beta = \beta({\Phi})
\end{eqnarray}

Let to introduce the new potential $\psi = \psi({\rho},{\Phi})$
according to

\begin{eqnarray}
\begin{array}{ccc}
\label{11}
q_{,1} &=& {b{\rho}}(\psi_{,2}+l\psi_{,1})e^{\alpha - \beta}/{{\Phi}^3},
\\ [4mm]
q_{,2}+lq_{,1} &=& {b{\rho}} \psi_{,1}e^{\beta-
\alpha}/{{\Phi}^3},\,\,\,\, b=Const.
\end{array}
\end{eqnarray}

Then, for the vector of the frame reference angular velocity

$$\omega^{\alpha} \equiv
-\frac{1}{2}{\varepsilon}^{{\alpha}{\beta}{\gamma}{\delta}}
{\left ( u_{[\beta,\gamma]}+u_{[\beta}{ F_{\gamma]}} \right )}
u_{\delta}$$
we obtain

\begin{eqnarray}
\label{12}
\omega_A= \frac{b\psi_{,A}}{2\Phi^2},\,\,\,A=1,2.
\end{eqnarray}

(Here $(...)_{,1}=\frac{\partial{(...)}}{\partial{\rho}},\,\,\,
(...)_{,2}=\frac{\partial{(...)}}{\partial{\Phi}}$).
Under these conditions the one from Einsteins equations turns identity and

$$ \frac{b\psi_{,1}}{\Phi^3}= \frac{\partial \omega_2}{\partial \rho}
 - \frac{\partial \omega_1}{\partial \Phi}. $$

Our basic supposition is following:the field of $\omega^\alpha$ vector
is irrotational i.e.

\begin{eqnarray}
\begin{array}{ccc}
\label{13}
\psi_{,1}=0,\,\,\,\omega={be^{-\beta}}\psi_{,2}/{2\Phi^2}
\\ [4mm]
q_{,2}+lq_{,1}=0.
\end{array}
\end{eqnarray}

Other Einstein equation give us
$$ \alpha_{,2}+l\alpha_{,1}=0. $$

In order to write the system of Einsteins equations of the considered
 problem it is necessary,taking into account given above conditions,
to find the first integral of the part of field equations and also
keep in mind

    i. the condition of the regularity on the rotation axis

$$\frac{e^{-2\alpha}{(\rho-qq_{,1}\Phi^2)}^2+e^{-2\beta}{(l\rho-q^2\Phi
)}^2}
{\rho^2-q^2\Phi^2} \stackrel{\rho \to 0} {\longrightarrow} 1.$$

   ii. the condition of the constancy the angular velocity

 $$ \frac{g_{t\varphi}}{g_{\varphi \varphi}}=
\frac{q\Phi^2}{\rho^2-q^2\Phi^2}= Const.,$$
so that  $q{(\rho \to 0) \to \rho^2}.$

Now the solution of the stationary axisymmetric GR problem in the metric
(6) and the co-moving frame of reference determined the following system
of the first order ordinary differential equations

\begin{eqnarray}
\label{14}
\beta_{,2}=L({\Phi})+2{\Phi}e^{2\beta}[\omega^2-2\pi{(\varepsilon+3\sl{P})}]
\end{eqnarray}

\begin{equation}
\begin{array}{ccc}
\label{15}
L_{,2}-L{\left \{ L/2+1/{\Phi}+2{\Phi}e^{2\beta}
[\omega^2-2\pi{(\varepsilon+3\sl{P})}] \right \} } =
\\ [4mm]
2e^{2\beta}[\omega^2-4\pi{(\varepsilon+\sl{P})}]
\end{array}
\end{equation}
\begin{equation}
\label{16}
\omega_{,2}/{\omega}+L-1/{\Phi}=0
\end{equation}
\begin{equation}
\label{17}
q_{,1}=-2\rho e^{\alpha} \omega/{\Phi}, \qquad q_{,2}+lq_{,1}=0,
\end{equation}
\begin{equation}
\label{18}
\frac{d \sl{P}}{d \Phi}=-(\varepsilon+\sl{P})/{\Phi}
\end{equation}
as well the algebraic relations

\begin{eqnarray}
\begin{array}{ccc}
\label{19}
e^{-2\alpha}=1-\rho^2 f(\Phi), \,\,\, l= \frac{1}{2}\rho L(\Phi)
\\ [4mm]
f(\Phi)=-8\pi \sl{P}-\omega^2 + Le^{-2\beta} {(4+L\Phi)}/{4\Phi}.
\end{array}
\end{eqnarray}

\section{ Vacuum solutions.}
In  vacuum case $(\varepsilon=\sl{P}=0)$
after the substitution $ y= {L \sqrt{a^4-b^2\Phi^4}}/{\Phi} $ the
key equation (15) amounts to

$$ y_{,2}= \frac{\Phi(y^2+4b^2)}{2 \sqrt{a^4-b^2 \Phi^4}} $$

and for $ L=L(\Phi) $ we obtain

\begin{eqnarray}
\label{20}
L= \frac{2\Phi}{\sqrt{a^4-b^2 \Phi^4}}
\frac{a^2(1-k^2 b^2)+(1+k^2b^2){\sqrt{a^4-b^2 \Phi^4}}}
{2ka^2-(1+k^2b^2) \phi^2},
\end{eqnarray}

where $a$ and $k$ are new constants. Using this expression, equations
(14) and (17) easily to find

\begin{eqnarray}
\label{21}
e^ \beta= 2Ca^2 \frac{1+k^2b^2}{\sqrt{a^4-b^2 \Phi^4}}
\frac{(1-k^2b^2){\sqrt{a^4-b^2 \Phi^4}}+a^2(1+k^2b^2)-2kb^2 \Phi^2}
{{[2ka^2-(1+k^2b^2) \Phi^2]}^2},
\end{eqnarray}

\begin{eqnarray}
\label{22}
q=\frac{bC(1+k^2b^2)}{ka^2}\left(e^{-\alpha}-1\right).
\end{eqnarray}

Let us to introduce the new variable $R$ in accordance with
\begin{eqnarray}
\label{23}
d\rho -ld\Phi = \rho d(lnR),
\end{eqnarray}
then
\begin{eqnarray}
\label{24}
\rho^2= \frac{R^2e^\beta\sqrt{a^4-b^2 \Phi^4}}{4Ca^2(1+k^2b^2)},
\end{eqnarray}
and the solution of vacuum problem acquired the form
\begin{eqnarray}
\label{25}
ds^2 = \Phi^2{(dt-qd\varphi)}^2-e^{2\beta}d\Phi^2-
\frac{\rho^2}{R^2}  \left[ { \left( 1-
\frac{kR^2}{2C^2a^2(1+k^2b^2)} \right)}^{-1}dR^2+R^2d\varphi^2 \right] .
\end{eqnarray}
Where

\begin{eqnarray}
\label{26}
q=\frac{bC(1+k^2b^2)}{ka^2}
\left({\sqrt{1-\frac{kR^2}{2C^2a^2(1+k^2b^2)}}-1}\right),
\end{eqnarray}

\begin{eqnarray}
\label{27}
e^\beta=\frac{C(1+k^2b^2)}{2ka^2}\sqrt{a^4-b^2 \Phi^4}
\left[{\frac{L^2}{4}+\frac{L}{\Phi}-\frac{b^2\Phi^2}{a^4-b^2 \Phi^4}}
\right]
\end{eqnarray}

Expressions (25)-(27) are the new solution of the stationary axisymmetric
GR vacuum problem

\section{ Static cases $(b=0)$.}

a) Let us assume $2ka^2=1$ and introduce coordinates $r$ and $\vartheta$
by
\begin{eqnarray}
\label{28}
\Phi^2=1-\frac{2M}{r},\,\,\,R=2M\sin{\vartheta},\,\,\,M=Ca^2.
\end{eqnarray}

In this case expressions (25)-(27) will be coincides with Shwarzshild
solution.

b) Let us assume $2ka^2=0$ and introduce $z=1/ \Phi^2$, then we obtain
\begin{eqnarray}
\label{29}
ds^2=\frac{dt^2}{z}-z^2(dR^2+R^2d\varphi^2)-4C^2a^4zdz^2,
\end{eqnarray}
or in equivalent form
\begin{eqnarray}
\begin{array}{ccc}
\label{30}
ds^2=e^{u/2Ca^2}dt^2-e^{-u/Ca^2}(dx^2+dy^2)-e^{-3u/2Ca^2}du^2,
\\ [4mm]
dx^2+dy^2=dR^2+R^2d\varphi^2,\,\,\,u=4Ca^2ln\Phi.
\end{array}
\end{eqnarray}

This solution coincides with Taub solution [4] for the gravitational
field of the flat uniform shell.

c) Let us assume $2ka^2=-1$ and introduce
$$ z=\frac{2Ca^2}{1+\Phi^2}, \,\,\,R=2Ca^2shr. $$

In this case we find
\begin{eqnarray}
\label{31}
ds^2=\left({\frac{2Ca^2}{z}-1}\right)dt^2-
\frac{dz^2}{({2Ca^2}/{z}-1)}-z^2(dr^2+sh^2rd\varphi^2).
\end{eqnarray}

These static special cases be the same as class A solutions for degenerate
vacuum static gravitational fields (cf [5]).

d) The incompressible fluid. Let  $p={\sl{P}}/\varepsilon_0$, where
$\varepsilon_0$=Const. The condition of the hydrostatic equilibrium
after integration give us
$$ 1+p=\frac{\Phi_s}{\Phi},\,\,\,p_s=\frac{\Phi_s}{\Phi_c}-1. $$
Here index "$c$" used for values in center of the mass distribution,and
index "$s$" for values on the boundary one. Let us introduce
$$ x=3\Phi_s-2\Phi,$$ and $$ u=e^{-2\beta}, $$
then from considered Einstein equation we obtain
$$ uu_{xx}=\frac{3}{4}{u_x}^2+4\pi\varepsilon_0 x(u_x+\pi\varepsilon_0 x)-
\frac{u{(u_x+4\pi\varepsilon_0 \Phi_s)}}{3\Phi_s-x}. $$

The partial solution of the last equation is
$$ u= \frac{2\pi}{3}\varepsilon_0(1-x^2). $$
therefore,
$$ e^{2\alpha}={\left[{1-\frac{8\pi\varepsilon_0}{3}
 \frac{\rho^2}{1-{(3\Phi_s-2\Phi)}^2}} \right]}^{-1},  $$

\begin{eqnarray}
\label{32}
e^{-2\beta}= \frac{2\pi}{3}\varepsilon_0
{\left[{1-{(3\Phi_s-2\Phi)}^2}\right]},
\end{eqnarray}
$$ l=\frac{2\rho (3\Phi_s-2\Phi)}{1-{(3\Phi_s-2\Phi)}^2}. $$

This solution may be rewrite using the Shwarzshild coordinates in the
following way
$$3\Phi_s-2\Phi=\sqrt{1-\frac{2M}{r}},\,\,\,\rho=r\sin{\vartheta},$$
\begin{eqnarray}
\label{33}
M= \frac{4\pi}{3}\varepsilon_0 {r_s}^3,
\end{eqnarray}
$$\Phi_s=1-\frac{2M}{r_s},\,\,\,
\Phi_c=\frac{3\Phi_s-1}{2}=\frac{\Phi_s}{1+p_c}.$$
\section{ The modified NUT solution.}
The well-known NUT [6] solution has
not the simple physical meaning. In our point of view this fact reasons
are the following

i. The NUT solution is not regularity along the symmetry axis

ii. This solution is not asymptotically flat.

Let us to consider the special stationary case of the solution obtained in
section 2, namely let

$$ 2ka^2=1+k^2b^2,\,\,\,\,\,M^2=C^2(a^4-b^2), \,\,\,\,\,n=Cb, $$
and introduced coordinates $r$ and $\vartheta$ as well
$$\Phi^2 = \frac{r^2-2Mr-n^2}{r^2+n^2},$$
\begin{eqnarray}
\label{34}
\rho^2 = (r^2+n^2){\sin}^2{\vartheta},
\end{eqnarray}
$$ r = \frac{M+\sqrt{M^2+n^2(1-\Phi^4)}}{1-\Phi^2}.$$

Then,

$$ ds^2 = \left({\frac{r^2-2Mr-n^2}{r^2+n^2}}\right){(dt+qd\varphi)}^2-
\left({\frac{r^2+n^2}{r^2-2Mr-n^2}}\right) dr^2- $$
\begin{eqnarray}
\label{35}
(r^2+n^2) \left({ {d\vartheta}^2+{\sin}^2{\vartheta}{d\varphi}^2 }\right)
\end{eqnarray}
$$ q =  2n(1-|\cos{\vartheta}|).$$

This solution different from NUT solution only   the expression for $q$
and can be reduced in asymptotically flat form after change the frame of
reference by transformation $ d\varphi=d{\tilde\varphi}+\omega dt$.
     In other special case  $2ka^2=1,\,\,Ca^2=A,\,\,b/a^2=B$ the general
solution (25)-(27) give us

$$ e^{2\alpha}={[1-\rho^2f(\Phi)]}^{-1} $$
$$ q=2AB{\left({1+\frac{B^2}{4}}\right)}\sqrt{1-B^2\Phi^4} $$
$$ e^\beta=A{\left({1+\frac{B^2}{4}}\right)}\sqrt{1-B^2\Phi^4}
\left[{\frac{L^2}{4}+\frac{L}{\Phi}-\frac{B^2\Phi^2}{1+B^2\Phi^4}} \right]
$$

\begin{eqnarray}
\label{36}
l=\frac{1}{2}\rho L(\Phi)
\end{eqnarray}
$$
f(\Phi)={\left[{Ae^\beta\left({1+\frac{B^2}{4}}\right)\sqrt{1-B^2\Phi^4}}\right]}^{-1}$$
$$
L(\Phi)=\frac{2\Phi}{\sqrt{1-B^2\Phi^4}}\frac{\left({1-\frac{B^2}{4}}\right)+
\left({1+\frac{B^2}{4}}\right)\sqrt{1-B^2\Phi^4}}
{1-\left({1+\frac{B^2}{4}}\right)\Phi^2} $$

Let rewrite the metric form (6) like following
\begin{eqnarray}
\begin{array}{ccc}
\label{37}
ds^2 = \left[{\rho^2\Phi^2}/{(\rho^2-q^2\Phi^2)}\right]dt^2-
e^{2\alpha}{(d\rho -ld\Phi)}^2-e^{2\beta}{d\Phi}^2-
(\rho^2-q^2\Phi^2){(d\varphi+\omega dt)}^2
\\ [4mm]
\omega={q\Phi^2}/{(\rho^2-q^2\Phi^2)}
\end{array}
\end{eqnarray}

If introduced
$$ \Phi^2=1-\frac{2Mr}{R^2} $$
$$ \rho^2= {\sin}^2{\vartheta}{\left( {r^2+a^2-
\frac{2Mar}{R^2-2Mr}+\frac{2Ma^2r{\sin}^2{\vartheta}}{R^2}} \right) } $$
$$ R^2=r^2+a^2{\cos}^2{\vartheta} $$
and transformed expression (37) according to

$$ d{\tilde \varphi}=d\varphi+{(\omega-\omega_k)}dt $$
$$ \omega_k=\frac{2aMr}{(r^2+a^2)R^2+2Ma^2r{\sin}^2{\vartheta}} $$

Then the special solution (36) coincides with the well-known form of
Kerr solution.

Thus the stationary axisymmetric problem is formulated using the
new nontraditional approach.We obtain the new vacuum solution of this
problem.The special case of founded solution is the modified NUT solution.
Other special case after transformations coincides with Kerr solution.

\end{document}